\documentstyle[prl,aps]{revtex}
\input{psfig.sty}
\draft
\begin{document}
\twocolumn[\hsize\textwidth\columnwidth\hsize\csname@twocolumnfalse\endcsname
\title{Stick-slip friction and nucleation dynamics of ultra-thin liquid films }
\author{I.S. Aranson$^1$, L.S. Tsimring$^2$, and
V.M. Vinokur$^1$ }
\address{$^1$Argonne National Laboratory,
9700 South Cass Avenue, Argonne, IL 60439\\
$^2$ Institute for Nonlinear Science, University of California,
San Diego, La Jolla, CA 92093-0402}
\date{\today}
\maketitle
\begin{abstract}
We develop the
theory for stick-slip motion in ultra-thin liquid films confined between
two moving atomically-flat surfaces. Our model is based on hydrodynamic
equation for the flow coupled to the dynamic order parameter
field describing the ``shear melting and freezing'' of the confined fluid.
This model successfully accounts for observed phenomenology of friction in
ultra-thin films, including periodic and chaotic sequences of slips,
transitions  from stick-slip motion to steady sliding.
%This approach substantially extends exiting
%phenomenological theories of shear melting and boundary lubrication on
%nanometer scale.
\end{abstract}
\narrowtext
\vskip1pc]
 
The nature of sliding friction is a 
fundamental physical problem of prime practical
importance \cite{persson,meyer}.
While the possibility to create low-friction surfaces and lubricant
fluids has been ubiquitous for almost all engineering applications, it has 
become crucial for design of modern micro-miniature devices such as
information storage and micro-electro mechanical systems, where
low friction without stick-slip (or interrupted) motion is 
necessary.

 
Studies of friction between  atomically flat mica surfaces separated by
the ultra-thin layer of fluid lubricant have revealed a striking
phenomenon \cite{israel1}: in a certain range of experimental parameters
the fluid exhibited solid-like properties, in particular, a critical
yield stress leading to stick-slips similar to that in solid-on-solid
dry friction process \cite{dry_fr}.  This behavior was attributed to the
confinement-induced freezing of the lubricant and its recurring melting
due to the increasing shear stress:
as the fluid thickness is reduced to several molecular layers, it
freezes, but when the shear stress exceeds some critical value, it melts
(so-called ``shear melting'' effect).  This behavior was confirmed
by molecular dynamics simulations \cite{thompson,landman} that indicated
ordering of the fluid due to confinement by the walls.

A quest for the quantitative description of the stick-slip lubricant 
dynamics motivated several theoretical works\cite{braun,zhong,batista}. 
The important step has been made by Carlson and Batista \cite{batista} who
proposed phenomenological constitutive relation connecting the
frictional forces to velocity and coordinates via the order parameter-like
state variable reflecting the degree of melting.
This model successfully described some of the observed phenomenology of
the experiment \cite{israel1} and gave a new insight into the dynamics. Yet
many important questions including the very mechanism 
of the onset of the stick-slip remain unresolved.

In this Letter we develop a theory of a stick-slip motion in an
ultra-thin confined liquids based on the equation for the flow coupled
to the equation for the order parameter (OP) for the melting transition
in the presence of the shear stress. We propose that the shear melting
is controlled by the stress tensor rather than the sliding velocity as
assumed in \cite{batista}. Making use of the generalized Lindemann
criterion, we combine shear and thermodynamic melting within a unified
description. Using this approach we describe the onset of the
stick-slip motion as the function of the film thickness and determine
the dynamic phase diagram.  We demonstrate that random nucleation of
droplets of the fluid phase during the motion leads to irregular temporal
distribution of slip events.

{\it Model}. The flow of liquid lubricant has to satisfy 
the momentum conservation law:
       \begin{equation}
	   \rho_0 \frac{D v_i}{Dt } = \frac{ \partial \sigma_{ij} }{\partial x_j},
	   \label{mom}
       \end{equation}
where $v_i$ is a component of the fluid velocity, $\sigma_{ij}$ is the stress tensor,
$D/Dt=\partial_t+ {\bf v}\nabla $ is the material derivative, 
and $\rho_0$ is the density of
fluid. Assuming incompressibility we set $\rho_0=1$ and
$\mbox{div} {\bf v}=0$.
%We assume that the hydrodynamic velocity $\bf v$ is small
%and replace $D /Dt \to \partial_t$.
We further assume that the lubricant satisfies Maxwell-type
stress-stain relation:
       \begin{equation}
	   \partial_t \sigma_{ij}+ \eta \sigma_{ij} = \mu U_{ij}
	   \label{ssr}
       \end{equation}
where $U_{ij}=\partial v_i/\partial x_j+\partial v_j/\partial x_i$
is the shear strain
rate, $\mu$ is the shear modulus, and $\eta$ is the shear stress relaxation 
rate.
%This relation reproduces both elastic behavior of solids for fast shearing
%and viscous
%behavior of fluids at slow shearing.
Thus, the stress-strain relation includes both viscous flow and elastic
restoring forces.
The conventional shear viscosity is defined as $\nu=\mu/\eta$.
 
%Eq. (\ref{ssr}) is insufficient 
To describe the dynamic phase transition between solid and fluid states
we take into account that the stress relaxation rate $\eta$ is itself a
function of the physical state of the material quantified near the
melting transition by the OP  $\rho$   which  is defined in such a way
that $\rho=1$ corresponds to the solid state and $\rho=0$ to the liquid
state. Physical interpretation of the OP for various systems can be
different, but for crystalline solids $\rho$ can be related to the
dislocation density.  We restrict ourselves to the simplest dependence
of the stress relaxation rate on $\rho$: $\eta=\eta_0 (1-\rho)$,
$\eta_0=const$.  This choice assures that the Eq. (\ref{ssr}) gives the
standard Hook's law for the pure solid ($\rho=1$) and the standard
viscous stress-strain relation for the Newtonian fluid with $\rho=0$.
We assume that the OP obeys the standard Ginzburg-Landau equation

\begin{equation}
\tau_0 \partial_t \rho = l^2 \nabla^2 \rho -\rho (1-\rho) (\delta -\rho)
 \label{ope1}
\end{equation}
where $\tau_0$ and $l$ are the characteristic time and length 
correspondingly;
$l$ should be of
the order of the lattice constant $a$,
and the characteristic time can be expressed through the
 sound velocity $c_s$, $\tau_0 \approx l/c_s$.
Here $\delta$ is the control parameter proportional to the temperature $T$.
Since  the melting transition in the lubricating layer
occurs under the out-of-equilibrium conditions, it is
characterized by two
critical temperatures, $T_1$, corresponding to an absolute instability of
the
overcooled liquid and $T_2$, the stability limit of the solid
phase\cite{T2,T1}.
The conventional thermodynamic melting temperature, $T_m$, is
confined between these limits: $T_1<T_m<T_2$.
The parameter $\delta$ is naturally
expressed in the form
     \begin{equation}
	 \delta=(T-T_1)/(T_2-T_1)
	 \label{delta}
     \end{equation}
Now we have to relate the solid instability temperature $T_{2}$ to
the stress generated in the process of motion. To this end we notice
first that in the absence of dynamic shear deformations the
temperature $T_{2}^{\circ}$ can be estimated as
$ T_{2}^{\circ}=\tilde c^{2}_{\scriptscriptstyle L}\mu a^{3}$,
where $a$ is the lattice constant for the solid crystalline state, 
$\mu$ is the shear 
modulus (we write everything in a scalar form for simplicity), 
and $\tilde c^{2}_{\scriptscriptstyle L}$ is the
numerical factor (``Lindemann number").
Assuming independence of thermal fluctuations and shear, one can present a 
rms displacement field in a form:
      \begin{equation}
\langle u^{2}\rangle \simeq \frac{T}{\mu a}+\frac{{\sigma}^{2}}{\mu^{2}},
	  \label{displ}
      \end{equation}
where $\sigma\equiv \sigma_{xy}$ is the shear stress, and
the first term in the rhs stands for the thermal average
displacement, while the second term expresses the shear-induced
displacement field. In writing Eq. (\ref{displ}) we hypothesize
that the solid phase instability can stem not only from the thermal
fluctuations, but also from the shear, generated by the
mutual motion of the solid surfaces confining the lubricant.  At
zero physical temperature the instability can be caused by this shear
only, this concept of {\it shear-induced melting} generalizes the
hypothesis of the dynamic disorder-driven melting introduced in earlier
work\cite{kosh}. At the temperature $T_{2}$ the relation
$ u^{2}=\tilde c^{2}_{\scriptscriptstyle L}a^{2}$ holds, and we
immediately obtain from Eq.(\ref{displ})
      \begin{equation}
	  T_{2}=\tilde c^{2}_{\scriptscriptstyle L}\mu
	  a^{3}-\sigma^{2}a/\mu
	  =T_{2}^{\circ}-\sigma^{2}a/\mu
	  \label{shrm}
      \end{equation}
Substituting Eq. (\ref{shrm}) into the expression for the control parameter
$\delta$ Eq. (\ref{delta}) one derives in the first order
        \begin{equation}
	    \delta=\delta_0+\sigma^2/\sigma_0^2
	    \label{delta1}
	\end{equation}
where $\delta_0=(T-T_1)/(T_2^\circ-T_1)$ and
$\sigma_0=\sqrt{ \mu(T_2^\circ-T_1)/a \delta_0} $
has a meaning of the yield shear stress.

In the {\it  thin layer approximation} we assume
that the thickness of the lubricant layer $h$ is small and
neglect the dependence of the shear stress $\sigma$
on the transverse coordinate $z$. 
We can also neglect the dependence of $\sigma$ on the longitudinal
coordinate $x$ if the sample size $L$ is not very large. Indeed,
if the ``acoustic shear time'' $\tau_a=L/c_s$
is much smaller then any characteristic time scale of the problem
(e.g stick-slip time),
the spatial variation of the stress is  negligible, and 
the shear stress becomes the  function of time only\cite{meer}. In this approximation
 after
the integration over the area of the sample, Eq. (\ref{ssr}) gives:
        \begin{equation}
	    \sigma_t + \frac{\sigma \eta_0}{ Lh} \int_0^L \int _0^h (1-\rho) dx dz=
	    \frac{\mu V}{h},
	    \label{ssr1}
	\end{equation}
where $V$ is the
relative velocity of the upper plate with respect to the bottom.

Now we can further simplify the OP dynamics.
Since the walls favor the formation  of the solid, the boundary conditions
for OP read: $\rho(0)=\rho(h)=1$, and the bulk variation of the OP
are small as compared to $1$. Let us  seek the
solution in a form
       \begin{equation}
	   \rho(x,z,t)=1-A(x,t) \sin (\pi/h z)
	   \label{rho1}
       \end{equation}
where $A \ll 1 $
is the slowly varying amplitude. Substituting Eq. (\ref{rho1}) into
Eq. (\ref{ope1}) and making use of the standard orthogonality procedure
leads to
      \begin{equation}
	  A_t = A_{xx} + (\delta -1 -\frac{\pi^2}{h^2}) A +
	  \frac{8}{3 \pi} (2-\delta ) A^2 -\frac{3}{4} A^3
	  \label{A1}
      \end{equation}
where $\delta=\delta_0 + \sigma^2$, and variable were rescaled as
$x/l \to x, t/\tau \to t, \sigma_{xy}/\sigma_0 \to \sigma$.
Accordingly, Eq. (\ref{ssr1}) yields:
        \begin{equation}
	    \sigma_t + \frac{ 2 \eta_0}{\pi L} \sigma \int_0^L A(x) dx = 
	    \frac{v_0 }{h},
	    \label{ssr2}
	\end{equation}
where $v_0= \mu V/\sigma_0 $ is the normalized pulling velocity.
%Eqs. (\ref{A1},\ref{ssr2}) are principle set of equations we will study.
Note that in the most of the dynamic friction
experiments the pulling velocity $\bar V$ does not
coincide with the upper plate velocity $V$. In fact,  the 
relation between the position of the upper plate $x$, friction force $F$, and
the position of the spring (neglecting the mass of the spring and
upper plate) is as follows: $F=k (\bar V t -X )$, where $X$ is the
horizontal position of the upper plate and $k$ is the stiffness of the spring.
Since $F \sim \sigma L$, and $V=dX/dt$, one immediately sees that exclusion of
the plate velocity $V$ 
results in the  equation that is qualitatively the same as
Eq. (\ref{ssr2}) but has renormalized $v_0$ and $\eta_0$. Thus to 
simplify further discussion  we will consider $V=\bar V$.

{\it Stick-slip motion}. First we discuss the spatially-uniform
motion.  In this case  Eqs.(\ref{A1}),(\ref{ssr2}) become
a pair of coupled ordinary differential equations (ODE).
Above the melting temperature $\delta_0>1$,
 the solid phase of the lubricant is formed due to
proximity-to-the walls effect.
In the experimentally relevant limit
$\eta_0, v_0 \ll 1$ the abovementioned ODEs can be investigated analytically 
by the
bifurcation analysis and the multi-scale technique with $\sigma$ 
being a slow
 and $A$ being a fast variable. Stick-slips are described by
the limit cycle on the $\sigma-A$ plane. The transitions between
different regimes are determined by the intersection of the manifold of 
the
slow motion $\frac{ 2 \eta_0}{\pi} \sigma A = v_0/h$ and the fast motion
$(\delta -1 -\frac{\pi^2}{h^2}) +
\frac{8}{3 \pi} (2-\delta ) A -\frac{3}{4} A^2=0$.
The transition from the stick-slips to sliding corresponds to 
the intersection of the
slow motion manifold with the minimum of the fast motion manifold
$\sigma=f(A)$, i.e. $d \sigma/ d A=0$. The 
limit cycle vanishes smoothly at the transition point if there is only
one intersection of two manifolds, otherwise the cycle
disappears abruptly with hysteresis.
%Typical diagram of regimes is shown in Fig. \ref{Fig1}.
%The phase diagram on $\delta_0,h$ plane is shown in in Fig. \ref{phase}.

\begin{figure}[h]
\centerline{ \psfig{figure=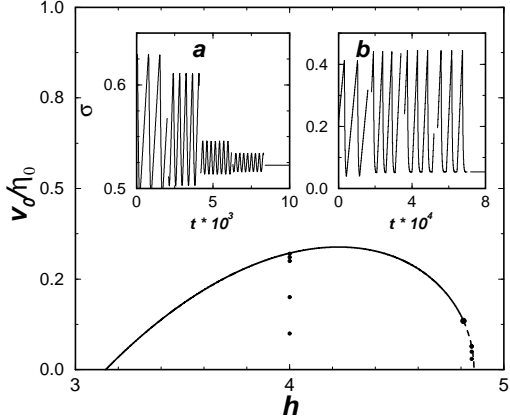,height=2.5in}}
\caption{Phase of lubrication regimes at $\delta_0=1.3$. 
Solid line indicates the continuous
transition from stick-slip to sliding, and
the dashed line corresponds to the hysteretic abrupt transition. 
Insets A,B show normalized shear stress $\sigma$ vs time
for $\eta_0=0.01$ at two different values of $h$ corresponding to
continuous and discontinuous transitions, and several values of 
$v_0/\eta_0$ approaching the transition line (dots in the main plot):
$h=4,\ v_0/\eta_0=0.1,0.2, 0.3, 0.31, 0.32 $ (A) and
$h=4.85\ v_0/\eta_0=0.03,0.05, 0.064, 0.0645, 0.065$ (B).
}
\label{Fig1}
\end{figure}
 
\begin{figure}[h]
\centerline{ \psfig{figure=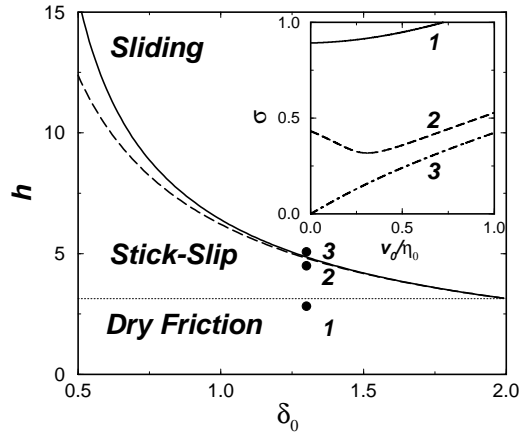,height=2.5in}}
\caption{Temperature ($\delta_0$)-thickness ($h$) diagram. 
Above solid line ($h_c$) sliding occurs for
arbitrary small velocity $v_0$. Left of $h_c$ stick-slips
exist for $v_0<v_c$, while the transition to sliding
is abrupt with hysteresis between
solid and dashed lines and smooth otherwise. Below dotted line
($h_0$) one has dry friction without stick-slips. Inset:
shear stress $\sigma$ (or friction force) vs sliding velocity $v_0$
in three different regions.
}
\label{phase}
\end{figure}

Figures \ref{Fig1},\ref{phase} illustrate the transition from
continuous sliding to the stick-slip motion.
As one can see from Fig. \ref{Fig1}, \ref{phase},
stick-slips are possible only in relatively thin layers, while in the thick 
layers sliding
is steady  since the lubricant in the bulk
is always in a fluid state. The critical thickness $h_c$ is determined 
by the
stability condition $d \sigma/ d A=0$:
$h_c=\pi /\sqrt{\delta_0-1+64 (2-\delta_0)^2 /27 \pi^2}$.
We find that for $h$ close to $h_c$
the transition from the stick-slip to sliding is always abrupt and has 
a hysteretic character (see Fig. \ref{phase} and Fig. \ref{Fig1}, Inset B).
For the parameters chosen, the ``friction law'' $\sigma$ vs $v_0$ has a minimum,
(Fig. \ref{phase}, inset, curve 2) as common for a typical
dry friction behavior\cite{persson}. 
For smaller
$\delta_0$ the  transition is continuous (inset A in Fig. \ref{Fig1},
curve 3 in Fig. \ref{phase}, inset).
For $h<h_0=\pi$ the dry friction without stick slip occurs because
the viscous friction force become larger then the dry friction one,
curve 1 in Inset \cite{note}. 

 

{\it Nucleation}. The above analysis presumed that the stick-slip
happens simultaneously in the entire space, which is certainly not 
the case for large samples.  It is natural to expect
that the stick-slip occurs via series of nucleation events when 
droplets of the liquid
emerge in the solid phase and then expand and merge throughout the system.
%Nucleation can be responsible for observed chaotic stick-slip
%sequences.
% (our second order ODE always give periodic oscillations).
It is interesting to connect the nucleation dynamics
in the systems with the lubricated friction  with 
the creation of topological
defects during rapid quench (``cosmological scenario'')\cite{kibble,zurek,akv}.
Since in the stick phase the value of $A$ is close to
zero, the solid phase can be significantly ``overheated" by the shear.
For $A \ll 1 $, integration of Eq. (\ref{A1}) yields
\begin{equation}
A\approx A_0 \exp\left(\int_0^t (-\epsilon + \sigma^2)
dt \right),
\label{A2}
\end{equation}
where $\epsilon=1 +\frac{\pi^2}{h^2}-\delta_0>0$.  From Eq.
(\ref{ssr2}), $\sigma\approx v_0 t/h + \sigma_0$, and $A_0, \sigma_0$
are the values of amplitude and stress in the beginning of stick phase.
We restrict ourselves to the case $h \to h_c$, where $\sigma_0 \to 0$
and $A_0 = O(1)$. While $\sigma$ is small, the amplitude $A$ decays
exponentially to very small values. It reaches a minimum
$A_{min}=\exp(-2\epsilon^{3/2}h/3v_0)$
at $t_{min}=\epsilon^{1/2}h/v_0$. Then it starts slowly to grow and
reaches the value $O(1)$
(slip event) at $t=t_m=\sqrt{3}t_{min}$. At $t_{min}<t<t_m$ the  growthrate 
$-\epsilon+\sigma^2>0$, and therefore the lubricant is in an unstable
(``overheated") state.  The overheated solid is very sensitive to
fluctuations, e.g. to thermal noise.  Small nuclei of liquid  can appear
and expand within the solid, resulting to an accelerated slip event.
Since at the low noise level nucleation events have the probabilistic
character, one can expect the spatio-temporal randomness of the slip
events.  At the higher level of the noise the slips become more
regular because the number of nucleation sites increases and the
overall effect of the noise is averaged out.

\begin{figure}[h]
\centerline{ \psfig{figure=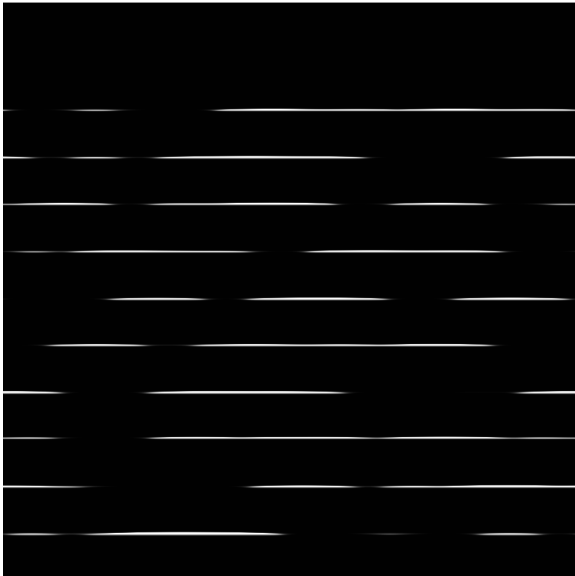,height=3.in}}
\caption{Space-time plot of $A$ for
$\delta_0=1.1$, $h=4$, $\eta=0.01$, $v_0=0.0002$ in the system of
length $L=1000$. Black correspond to $A=0$(solid), white to $A=1$ (liquid).
Time progresses from top to bottom, total integration time
$30000$ dimensionless units. Uncorrelated noise with zero average
and amplitude $10^{-16}$ is add on each time step and each grid point.
}
\label{Fig2}
\end{figure}

\begin{figure}[h]
\centerline{ \psfig{figure=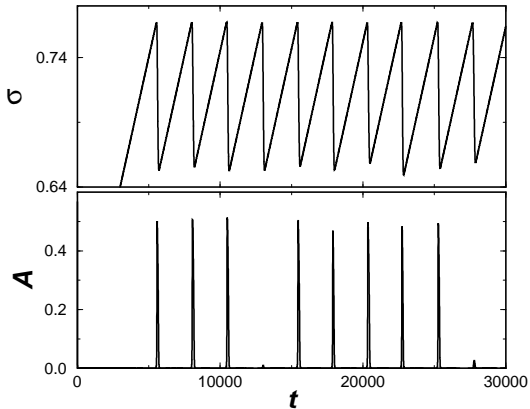,height=2.5in}}
\caption{$\sigma$ and $A$ at $x=L/2$ vs $t$
for parameters of Fig. \protect \ref{Fig2}.
}
\label{Fig3}
\end{figure}

We studied Eqs.
(\ref{A1},\ref{ssr2}) numerically in a fairly large domain, see
Figs. \ref{Fig2},\ref{Fig3}.
Since during the slip phase the shear stress
rapidly drops, the domains do not necessary
propagate through the entire system and in large systems
one may observe ``partial slips.''
The random character of the nucleation process manifests itself in the
non-periodicity of slips and local amplitude 
$A$.
 

{\it In conclusion}, we have investigated the friction dynamics in 
ultra-thin liquid films.  We have related the stick-slip
behavior of thin liquid films to shear-induced melting and freezing 
processes.  The developed approach allowed for the first time a 
quantitative description of dynamic nucleation effects leading to slip
events.  The proposed theory can be applied to a wide range of
phenomena including friction in nanoscale devices, friction on ice,
granular materials \cite{gollub}  as
well as depinning transitions of flux-line lattices in
type-II superconductors, charge density waves and other structures 
driven through disorder \cite{kosh}. Our model offers a description for the
long-standing problem of the  ultra-sound emition during the friction 
dynamics.
This research is supported by the Office of the
Basic Energy Sciences at the US DOE, grants W-31-109-ENG-38,
DE-FG03-95ER14516, and DE-FG03-96ER14592


\vspace{-1cm}
 
\references
\bibitem{persson}
B.N.J. Persson, {\it Sliding Friction. Physical Principles and Applications},
Springer-Verlag, 1998
\bibitem{meyer}E. Meyer, R.M. Overney, K. Dransfeld, and T. Gyalog,
{\it Nanoscience. Friction and Rheology on the Nanometer Scale},
World Scientific, 1998
\bibitem{israel1} H. Yoshizava and J. Israelachvili,
J. Phys. Chem. {\bf 97}, 4128 (1993)

\bibitem{dry_fr} 
G. H\"ahner and  N. Spencer, Phys. Today {\bf 22} (9), 22 (1998); 
T. Baumberger, P. Berthoud, and C. Caroli, \prb {\bf 60}, 3928 (1999); 
 

\bibitem{thompson} P.A. Thompson and M.O. Robbins, \pra {\bf 41}, 6830 (1990);
Science {\bf 250}, 792 (1990)

\bibitem{landman} J.P. Gao, W.D. Luedtke, and U. Landman, J. Chem. Phys.
{\bf 106}, 4309 (1997)
\bibitem{braun} O. Braun, A.R. Bishop and J. R\"oder, \prl {\bf 82}, 3097 (1999)
\bibitem{zhong} W. Zhong and D. Tomanek, Europhys. Lett. {\bf 15}, 887 (1991)
\bibitem{batista} J.M. Carlson and A.A. Batista, \pre {\bf 53}, 4153 (1996).
\bibitem{T2}L. Pietronero, in {\it Phonons: Theory and Experiments},
ed. P.Br\"uesch (Springer, Berlin, 1987), Vol. III, Chap. 8.
\bibitem{T1}T.V. Ramakrishnan and M. Yousouff, Phys. Rev. B {\bf 19},
2775 (1979)
\bibitem{kosh}A. E. Koshelev and V. M. Vinokur, Phys. Rev. Lett., {\bf 73}, 3580 (1994)
\bibitem{note}
Curves 1 and 2 of Fig. \ref{phase} 
 are reminiscent of the  
generic  behavior of lattices driven
through random environment, near the depinning transition:  
curve 1 describes {\it elastic} depinning from weak disorder 
and  curve 2 represents hysteretic depinning corresponding to
the case of strong disorder, see e.g. 
V. M. Vinokur and T. Nattermann,  \prl {\bf 79}, 3471 (1997); 
M. C. Marchetti, A. Middleton, and T. Prellberg,  
\prl {\bf 85}, 1104 (2000)
\bibitem{meer} Validity of similar condition of isobaricity
in the context of radiative condensation is discussed in detail
by B. Meerson, \rmp {\bf 68}, 215 (1996), and
I. Aranson, B. Meerson, P. Sasorov, \pre
{\bf 52}, 948 (1995).
\bibitem{kibble} T.W.B. Kibble, J. Phys. A: Math Gen {\bf 9}, 1387 (1976)
\bibitem{zurek} W. H. Zurek, Nature {\bf 317}, 505 (1985);
N.D. Antunes, L.M.A. Bettencourt, and W.H. Zurek,
\prl {\bf 82}, 2824, (1999).
\bibitem{akv} I.S. Aranson, N.B. Kopnin and V.M. Vinokur, \prl {\bf 83},
2600 (1999)
\bibitem{gollub} W. Losert {\it et al}, \pre {\bf 61}, 4060 (2000)
\end{document}